\documentclass[journal=ancham,manuscript=article]{achemso}
\usepackage[utf8]{inputenc}
\usepackage{amsmath}
\usepackage{xcolor,booktabs}
\usepackage[normalem]{ulem}

\author{Yves Kayser}
\affiliation[PTB]{Physikalisch-Technische Bundesanstalt, Abbestraße 2-12, 10587 Berlin, Germany}
\email{yves.kayser@ptb.de}
\author{János Osán}
\affiliation[EK]{Environmental Physics Department, Centre for Energy Research, Konkoly-Thege M. út 29-33., 1121 Budapest, Hungary}
\author{Philipp Hönicke}
\author{Burkhard Beckhoff}
\affiliation[PTB]{Physikalisch-Technische Bundesanstalt, Abbestraße 2-12, 10587 Berlin, Germany}

\title{Reliable compositional analysis of airborne particulate matter beyond the quantification limits of total reflection X-ray fluorescence}


\begin{document}

\begin{abstract}
Knowledge on the temporal and size distribution of particulate matter (PM) in air as well as on its elemental composition is a key information for source appointment, for the investigation of their influence on environmental processes and for providing reliable data for climate models. While cascade impactors allow for time- and size-resolved collection of airborne PM, total reflection X-ray fluorescence (TXRF) allows for element-sensitive investigation of minute sample amounts thanks to its detection sensitivity. But during quantification by means of TXRF it is crucial to be aware of the linear calibration limits of TXRF in order to identify situations where collection times or pollution levels in the different size partitions were exceedingly long or high. Indeed, TXRF can only be reliably used when the amount of matter collected on the top of substrate is sufficiently dilute. By means of grazing incidence X-ray fluorescence (GIXRF), where the excitation conditions are varied in a controlled and reliable manner and include also the TXRF regime, a self consistent quantification of elemental mass depositions can be performed in order to validate or falsify TXRF quantification results. For low mass depositions an agreement within few percent for the different excitation conditions was found, while for increasing amounts of material relative errors of up to factor of 4 were found for TXRF as compared to GIXRF. Thus, TXRF cannot be applied to all samples regardless of their coverage and threshold values for the validity of quantification results need to be determined. As a flexible solution, GIXRF allows extending the dynamic range of reliably quantifiable mass depositions beyond the linear regime of TXRF, an important advantage when variable amounts of airborne PM need to be quantified as in the case of collection with cascade impactors. Furthermore, the presented more reliable quantification approach can be transferred to mobile tabletop instrumentation as well. This aspect is highly relevant for air quality monitoring in terms supporting the definition of appropriate legislation and measures for health and climate protection as well as for supporting their enforcement.
\end{abstract}

%
%
%

\section{Introduction}
Aerosols present in the environment affect our daily life at multiple levels. For example, airborne particulate matter (PM) in air can impact health due to inhalation \cite{Apte2015, Lelieveld2015, Leni2020} or can influence atmospheric processes \cite{Mellouki2015}, more precisely the climate and environmental ecosystems through impacting cloud formation \cite{Faye2015} or reflecting and scattering sunlight \cite{Zhang2015}. The chemical composition, which requires element-sensitive analytical methods, as well as the chemical speciation of the elements contained in airborne PM is of interest for a correct comprehension of its physical and chemical properties \cite{Prather2008}. With regard to health concerns, the smallest particles, so called fine and ultrafine particles with sizes in the sub-micrometer and sub-100 nanometer range, are the most concerning for epidemiology as they can penetrate into the airways of the lungs and may be held responsible for health-averse effects on the respiratory and cardio-vascular system upon long-term exposure \cite{Moller2008,Stone2007,Vavlanidis2008,Brook2010}. In particular anthropogenic emissions result in a noticeably higher generation of ultrafine particles \cite{Ronkko2019}. While toxicological studies have to assess possible health risks of different nanomaterials \cite{Daellenbach2020}, parallel efforts have to be undertaken to quantify the presence of the different elements in air and trace back the physical processes airborne PM is undergoing under different weather conditions. The compositional analysis of aerosols is often addressed by regulated analytical techniques requiring moderate amounts of substance collected in fiber filters. Improved time-resolved and size-fractionated information is, however, relevant for accurate modeling of climate changes, for regulatory bodies to impose preventive measures and for legal entities to enforce regulations on air quality and to correctly pinpoint anthropogenic or natural sources. Adding to this the requirement to not only detect but to quantify reliably trace levels of aerosols contained in air in order to achieve good time resolution during environmental monitoring campaigns, highly sensitive and accurate techniques need to be used.

Among  different available techniques \cite{Bulska2017,Ault2017} X-ray fluorescence (XRF) based methods are promising contributors to the field by delivering ensemble information on the chemical composition due to the advantages provided in terms of sample preparation, consumption and sensitivity \cite{Bulska2017,Furger2017}. The best possible detection limits with XRF based techniques can be achieved by means of total reflection XRF (TXRF) \cite{Beckhoff2007,Streli2008}. In combination with cascade impactors, where the PM is collected on different impaction stages and discriminated by their size using the principle on inertia, all relevant information for a time-dependent, element- and size-sensitive quantification of airborne PM in a defined volume of air down to the range of few ng/m$^3$ are at hand \cite{BUKOWIECKI2008,Richard2010, Bontempi2010, Prost2017,Osan2020,BORGESE2020, Fomba2020}. But varying environmental conditions and human activities result in unpredictable pollution levels which may in addition severely vary between the different impaction stages for each collection interval. Thus, it can not be ensured beforehand that the samples from the different impaction stages and for the different collection times are all within the linear regime of TXRF where it is assumed that the X-ray standing wavefield (XSW) created is not or only to a small amount disturbed by the material on the top of the substrate. Since the impact of airborne PM on the XSW can not be known beforehand, quantification by means of TXRF, which is a single-point measurement technique without parameter variation, risks to fail and must be assessed during outdoor campaigns. 

In the present work, the impact of different mass depositions on the reliability of the quantification of the elemental mass deposition by means of TXRF is investigated. This assessment is being done by analyzing samples from field campaigns with different collection times by both, standard TXRF at a single, fixed incidence angle and by grazing incidence X-ray fluorescence (GIXRF) where the incidence angle is varied in a controlled and reproducible manner. Indeed, during a GIXRF measurement the X-ray standing wavefield (XSW) created on the top of the sample is modified, respectively can be neglected for incidence angles far above the critical angle of total external reflection such that the excitation conditions are fundamentally modified during the measurement. For any given sample the mass deposition of the airborne PM collected depends on the pollution level and collection time and does not vary during analysis. Hence, it must be expected that the quantification of the different elements yields the same result for each incidence angle used during the GIXRF measurement. If this is the case, the quantification result can be considered as validated in a self-consistent approach. Hereafter, it will be assessed whether quantification of elemental mass depositions using TXRF and a self-consistent validation of the results can be achieved independently of the amount of airborne PM collected. 

\section{Quantification by means of TXRF}
TXRF employs a specific geometry in which the X-ray beam used for the excitation of the XRF signal impinges the sample at a very shallow angle beneath the critical angle for total external reflection \cite{Klockenkamper2014book}. Hence, TXRF demands for collimated and monochromatic excitation conditions but offers advantages such as the illumination of large sample areas and a large solid angle of detection. Further benefits offered by TXRF are twofold. On the one side, the penetration of the incident beam into the substrate is reduced to an evanescent wave and any background signal, XRF or scattering, originating from it is suppressed. On the other side, the reflection of the incident X-ray beam at the substrate surface leads to the creation of a XSW due to interference between incident and reflected X-rays. Consequently, enhanced excitation conditions for XRF originating from the particulate matter deposited on the top of the substrate can be achieved. A prerequisite to profit from total external reflection is to use substrates which are flat on a macroscopic scale and characterized on a microscopic scale by a roughness smaller than the wavelength of the incident X-rays for best possible reflectivity.

In the field of environmental sciences, this necessity for TXRF measurements prohibits analyzing foils or membrane filters as it is realized with different instruments used for the collection of airborne PM. In this case, the sampled material needs first to be transferred onto an adequate substrate which is realized usually via different digestion techniques \cite{Leland1987,SCHMELING1997,WAGNER2010,Fomba2020} or slurry techniques \cite{Natali2016,Bilo2017}. However, this approach discards TXRF based analytics of its main advantages since such a time-consuming preparation step may involve sample digestion, material loss or contamination issues. A more suitable approach is to use the TXRF substrates directly, as can be done in cascade impactors, to collect the particulate matter, either as they are \cite{SCHNEIDER1989,ESAKA2003,Osan2020}  or after applying a coating (to prevent bounce-off effects for example) and removing it after the sample collection \cite{INJUK1995,SCHMELING1997,Prost2017,PROST2018}. This additional a posteriori treatment prior to TXRF investigation \cite{Prost2017} is detrimental if other, complementary analytical techniques shall be used as well. 

Regardless of the sample preparation, the knowledge of the XSW is of importance for quantitative measurements independently if external standardization, internal standardization or reference-free quantification schemes are applied. When using internal standardization the XSW created needs to be identical throughout the sample area illuminated, while in case of external standardization the same XSW needs to be created in a reproducible manner for all samples investigated. External standardization means that for each element of interest in an experimental campaign a calibration curve is established by means of a set of reference samples with different mass depositions of the selected elements. This approach requires adequate reference samples which are sufficiently representative of the samples investigated \cite{HONICKE2018,HORNTRICH2012}. For samples collected during outdoor campaigns the criteria include elemental composition (sample matrix), mass deposition (concentration), particle size range and morphology as well as deposition pattern. Hence, the production and selection of adequate calibration samples for outdoor sampling campaigns requires additional a priori information. The characterization of actual samples from the measurement campaign via complementary techniques in order to use these samples as a kind of standards is often impeded by the sensitivity, i.e., the amount of sample required, of these techniques. 

For this reason approaches based on internal standardization were developed as an alternative that can be used with digested samples or in conjunction with substrates prepared for sampling \cite{Richard2010,BOTTGER2018}. Internal standardization means that on each sample to be analyzed a known quantity of a reference element is added beforehand of the measurement while assuming that the excitation and detection conditions at the position where the standard is deposited is representative for the whole sample. However, the standard which is added needs to fulfill the requirements of non-toxicity, not being ubiquitous and having XRF energies which do not overlap with the XRF lines to be contained within the sample. The goal of both approaches, external and internal standardization, is to extract combined information on instrumental factors in order to allow quantifying the elemental content of the material deposited on the top of the substrates. 

In the reference-free XRF quantification scheme \cite{Beckhoff2007}, information on the different experimental and fundamental parameters is used to calculate the mass deposition of different elements from the measured count rate of the corresponding fluorescence line. This approach requires the use of (radiometrically) calibrated instrumentation, e.g. apertures for an accurate knowledge of the solid angle of detection as well as efficiency of diodes and the silicon drift detector (SDD) for the incident photon flux and the detected XRF intensity, and the knowledge of atomic fundamental parameters (FPs), i.e. photoionization cross-sections and fluorescence yields which are element-dependent and in a large part also energy-dependent. 

\section{Samples}
The size-fractionated sampling of PM was realized by means of a 9-stage extension of the May-type cascade impactor \cite{May1975}. The time-resolved aspect is subject to the collection time and should be kept within a suitable range of collected mass for TXRF analysis in order to best profit from the sensitivity offered by this technique and avoid biasing the quantification as discussed in this work. The aerodynamic cut-off diameters of the stages 1 to 9 are respectively 17.9 $\mu$m, 8.9 $\mu$m, 4.5 $\mu$m, 2.25 $\mu$m, 1.13 $\mu$m, 0.57 $\mu$m, 0.29 $\mu$m, 0.18 $\mu$m and 0.07 $\mu$m at a constant flow rate of 16.7 L$/$min. The cut-off diameter is defined as the dimension of the PM which is collected with 50$\%$ efficiency, smaller particles escaping with a higher probability. A well-known and constant airflow is required during the collection of airborne PM. The first two stages with the largest particle sizes are disregarded in general for X-ray analysis and for the 7 further stages $20 \times 20$ mm$^2$ Si wafers are used as substrates. As Si wafers have very low background contamination and very low surface roughness, they are ideally suited for TXRF and GIXRF experiments. Measurements with good signal-to-background ratio can be expected even though other substrates might be more suitable for the collection of PM. Indeed, the collection efficiency depends not only on the design of the airflow, where losses of particles should be minimized \cite{Marple2004}, but also on the substrate surface. In order to preserve the capabilities offered by TXRF, no pre-treatment of the Si wafers was used. Sample sets selected for the present study were collected from two campaigns at two cities, Budapest, Hungary, 24-31 May 2018; and Cassino, Central Italy, 20-27 September 2018, with sampling duration ranging from 20 min to 5 h. In total 19 Si substrates collected on the 3 stages with the finest particle distributions were used in this survey. Among these 19 samples 6 samples will be discussed in more details and further information on these samples is provided in Table \ref{tab:table1}. These 6 samples are representative for the range of deposited particulate mass quantified for all 19 samples.
A summary of the quantification results for all 19 samples for the different excitation regimes and elements in Fig. S1.

The deposition area from the May-type cascade impactor corresponds to a stripe of 20 mm length and, depending on the stage, of 0.1 to 1 mm width (fine to coarser PM). The width is determined by the width of the slits used as nozzles for the different stages. This type of deposition pattern presents the advantage of being highly suitable for investigation by means of TXRF and GIXRF once the stripe is aligned along the incidence direction. Thus, the May-type cascade impactor is ideal to demonstrate the capability offered by the combination of cascade impactors and TXRF, respectively GIXRF analysis to provide element, size- and time-resolved information on the PM collected.

\begin{table}[htb]
    \centering
		    \caption{Description of the 6 selected Si substrates with aerosol particles collected in Budapest and at Cassino which are discussed in more details in Figs. \ref{fig:figure2} and \ref{fig:figure4}. Samples are listed in the order of deposited particulate mass and cover the full range of elemental mass depositions quantified on the total of 19 samples investigated.}
    \label{tab:table1}
    \begin{tabular}{|c|c|c|c|}
 \hline
 Sample & Duration & Stage & Diameter range / nm \\
 \hline
A & 30 min & 7 & 300 - 600 \\
B &	20 min & 9 & 70 - 180 \\
C & 20 min & 8 & 180 - 300 \\
D &	1 h & 7 & 300 - 600 \\
E &	5 h & 9 & 70 - 180 \\
F &	5 h & 7 & 300 - 600 \\
 \hline
    \end{tabular}

\end{table}

\section{Experimental}
\begin{figure}
	\begin{center}
	\includegraphics[width=0.7\textwidth]{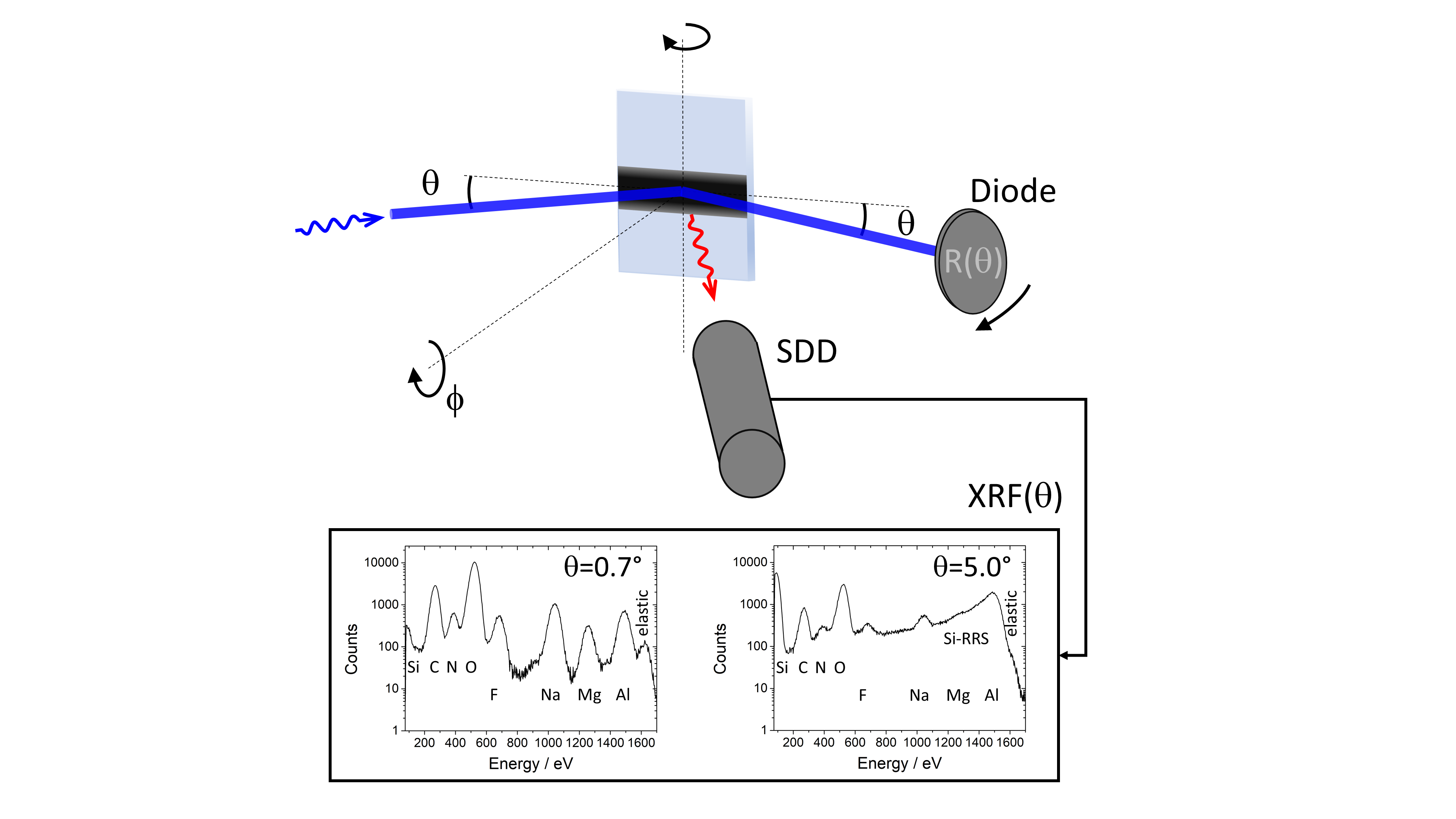}
	\end{center} 
	\caption{Illustration of the experimental setup with the Si wafer and the collected airborne PM on the top of it, a diode for measuring the reflectivity and a calibrated SDD for recording the XRF emitted for different incidence angles $\theta$ (top panel). Typical XRF spectra recorded for an incidence angle beneath and above the critical angle of total external reflection illustrate the lower background contributions from the Si wafer at the smaller incidence angle (bottom panels).}
	\label{fig:figure1}
\end{figure}

The reference-free GIXRF measurements \cite{Beckhoff2007,Muller2014,Hoenicke2019} for quantification of elemental mass depositions were realized at the plane grating monochromator (PGM) \cite{Senf1998} beamline in the PTB laboratory at the BESSY II electron storage ring. The experiments were conducted at an incident photon energy of 1620 eV which is below the Si K-edge in order to suppress the contribution of the Si K XRF lines. An ultrahigh-vacuum chamber equipped with a 9-axis manipulator was used \cite{Lubeck2013}. The use of an ultrahigh vacuum may induce loss of volatile material, for example organic compounds, but this was not further considered in this work where the focus lies on the evaluation of analytical techniques. The instrument allows for precisely tuning the incident angle $\theta$ between the incidence direction of the synchrotron radiation and the sample surface (Fig. \ref{fig:figure1}). The fluorescence radiation emitted from the sample was detected by means of a silicon drift detector (SDD) calibrated in terms of response function \cite{Scholze2009} and detection efficiency \cite{Scholze2005}, which is placed in the polarization plane and perpendicular to the propagation direction of the linearly polarized incident X-ray beam in order to minimize scattered radiation. The SDD allows for an energy-dispersive detection of the XRF emitted from the sample such that the information from different elements can be discriminated and processed in parallel during quantification. The incident photon flux is determined by using a calibrated photodiode. The spectra were deconvoluted using the known detector response functions for the relevant fluorescence lines and background contributions, which was mainly resonant Raman scattering (RRS) from the Si K shell \cite{Muller2006} and to a lower extent Bremsstrahlung from L shell electrons from the Si substrate. The resulting count rate $I$ for each fluorescence line of interest is normalized with respect to the sine of the incident angle $\theta$, the incident photon flux $I_0$, the effective solid angle of detection $\frac{\Omega(\theta)}{4\pi}$ and the energy dependent detection efficiency $\epsilon (E)$ of the SDD for the respective fluorescence photons in order to derive the emitted fluorescence intensity. It has to be emphasized that the calculation of the incident angle dependent solid angle of detection requires an accurate knowledge of the detection geometry but also of the incident beam profile. 

\begin{figure}
	\begin{center}
	\includegraphics[width=0.8\textwidth]{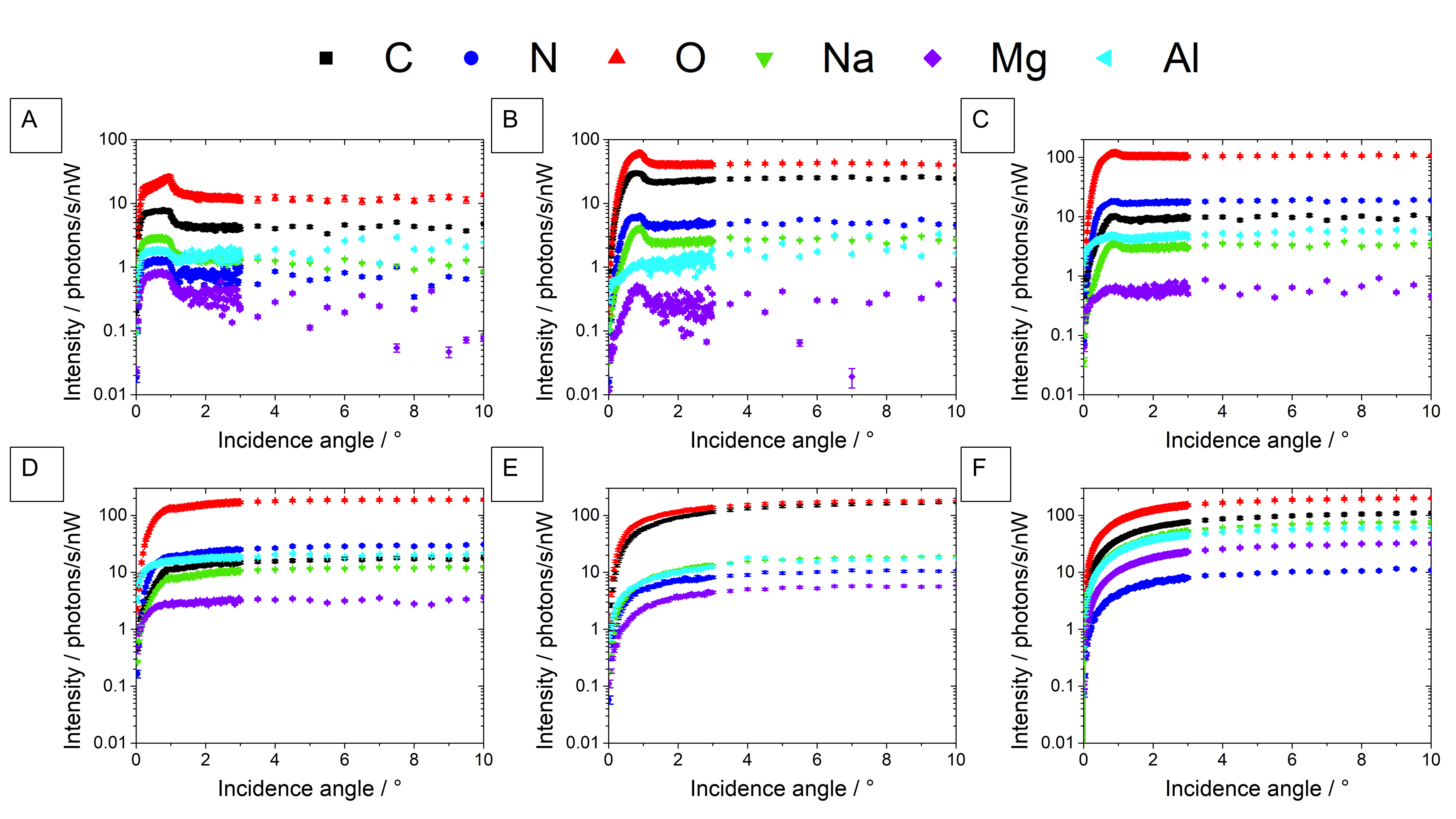}
	\end{center} 
	\caption{GIXRF data for the 6 selected different samples (labelled A to F and described in Table \ref{tab:table1}). The changes in the angular intensity profiles for each element indicate differences in the excitation of the XRF signal. The typical particle-like signature of the main elements detected in the GIXRF measurement gradually vanishes which is a clear indicator that the XSW on the top of the substrates significantly differs between the samples. It can be noted as well that for sample A, the angular evolution of O contains both particle- and layer-like signatures. The latter contribution arises from the surface oxide of the Si wafers used, but the relative contribution vanishes with increasing mass of collected airborne PM.}
	\label{fig:figure2}
\end{figure}

From the absolute XRF intensity the elemental mass deposition $m_{\text{A}}$, defined as mass per unit area, can be extracted for each position of the GIXRF measurements where the incident angle $\theta$ was varied in variable steps from 0$^\circ$ to 10$^\circ$ (Fig. \ref{fig:figure2}). Hence, the excitation conditions on each sample were gradually modified from total reflection conditions, where an XSW needs to be considered, to shallow incidence angle conditions, where no XSW is present. Nevertheless shallow incidence angles still provide efficient excitation conditions by dispersing the incident X-ray radiation over a larger sample area and increasing the incidence path length through the PM collected along the stripe-like deposition pattern. The calculation of the XSW requires the knowledge of the optical properties of the substrate, including possible surface oxidation, for the incident photon energy used during the experiment. However, even if the incident photon energy dependent optical properties are measured beforehand from a blank Si substrate to not rely on tabulated data, the presence of the PM on the top of the Si wafer will impact the reflectivity to a certain extent as the contrast in electronic density at the interface separating the bulk Si from the vacuum or PM is changing \cite{Unterumsberger2020}. Therefore, the reflectivity $R(\theta)$ for each wafer was measured by means of a photodiode positioned in a $\theta$ - $2\theta$ configuration during the GIXRF measurement. This approach is novel and allows for a reliable direct calculation of the incident angle dependent XSW for each sample under the actual measurement conditions. Any deviation from optimal alignment of the sample, increased surface roughness of the substrate or impact of the collected airborne PM on the XSW will be directly considered in this experimental approach. 

\section{Reference-free GIXRF Quantification}
\begin{figure}
	\begin{center}
	\includegraphics[width=0.4\textwidth]{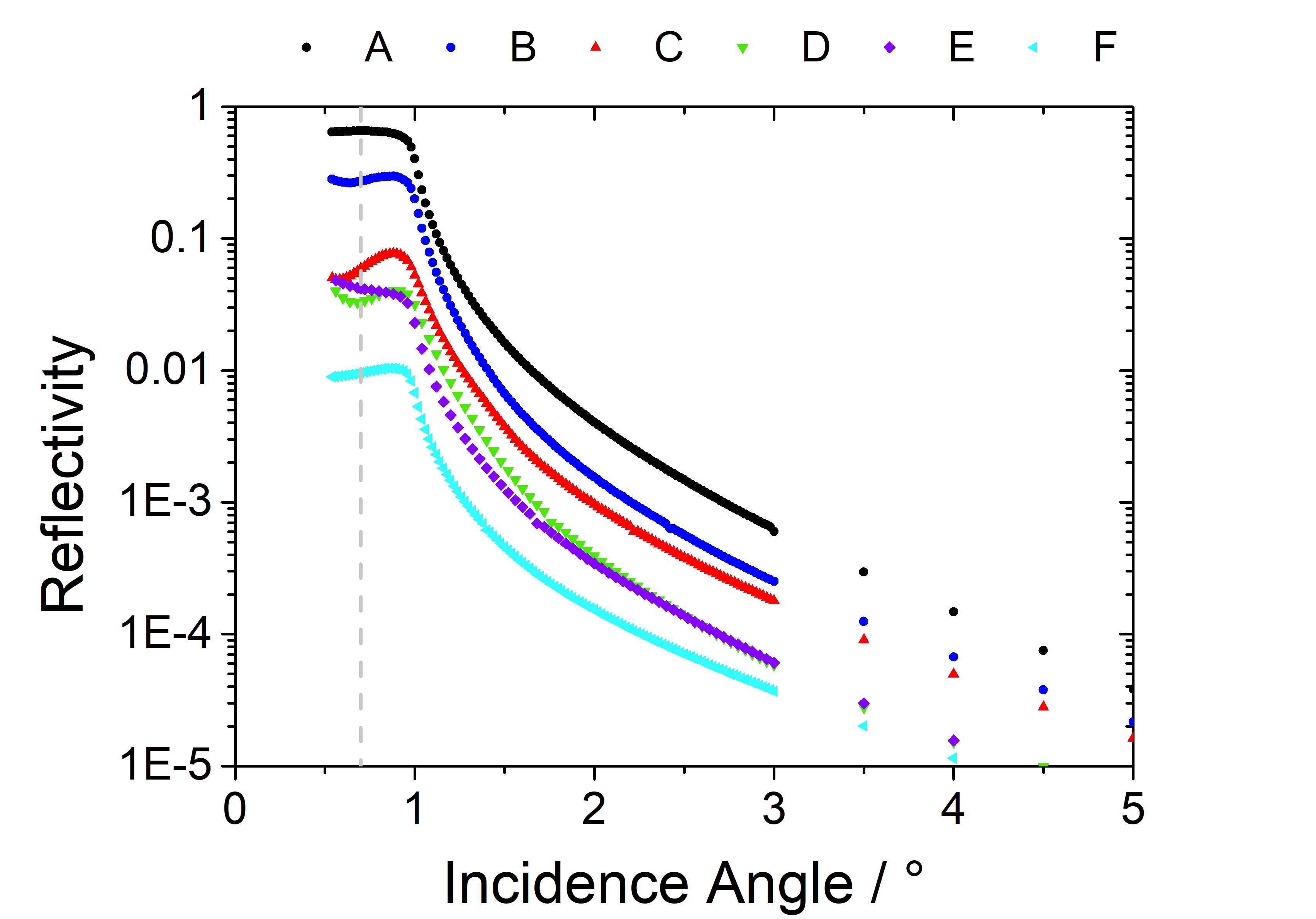}
	\end{center} 
	\caption{The reflectivity from the Si substrate for the same samples than displayed in Figs. 2 and 4 indicates the growing impact on the attenuation of X-rays within the collected PM, resulting in significant differences in the $\overline{XSW}(\theta)$ between the different samples. The vertical bar in the left panel indicates the position typically selected for a TXRF measurement for a Si substrate and the incident photon energy used. This position was used for the calculation of $\overline{XSW}(\theta)$ (Fig. S3).} 
	\label{fig:figure5}
\end{figure}

In a TXRF measurement the XRF intensity is usually recorded at a single incidence angle corresponding to $\frac{1}{\sqrt{2}}$ ($\approx 70\%$) of the critical angle for total external reflection $\theta_c$, which depends on the incident photon energy and the substrate density \cite{Klockenkamper2014book} and which was about 1$^\circ$ for the Si wafers used. The relative intensity distribution within the XSW is given by \cite{Klockenkamper2014book},
\begin{equation}
XSW(\theta,z)=1+R(\theta)+2\sqrt{R(\theta)}\,\cos\left(\arccos(2\frac{\theta^2}{\theta_c^2}-1)-4\pi\sin\theta\,\frac{z\,E_0}{hc}\right)
\end{equation}
with $E_0$ the energy of incident photons, $z$ the height above the reflecting substrate and $R(\theta)$ the measured reflectivity. Since the projection of the incident beam has to be below the sample dimension to discard scattering contributions to the signal, the reflectivity could only be accurately measured for incidence angles above 0.6$^\circ$. For smaller angles a geometrical correction needs to be considered with the consequence of a larger error in the calculation and hence the quantification result. As already noted, the use of experimental reflectivity data allows for more reliable calculations of the XSW  without requiring assumptions in how far the airborne PM collected will impact the XSW created on the top of the substrate. In Fig. \ref{fig:figure5} it is shown that the reflectivity at a typical angle used for TXRF measurements drops significantly with increasing mass deposition of airborne PM. This observation means that the XSW is significantly different for each sample and different compared to the case of a blank substrate (Fig. S2).

In the following the mean intensity of the XSW, labeled $\overline{XSW} (\theta)$, over the direction vertical $z$ (Fig. \ref{fig:figure1}) to the substrate surface is considered. The assumptions of a laterally and vertically homogeneous chemical composition of the collected PM and of PM dimensions extending over several periods $\frac{hc}{2\,E_0\,\sin\theta}$ of the XSW are made thereby. The averaging of the XSW by integration is further backed up by the fact that different particles sizes and compositions are intermixed on each stage and that the deposition pattern is homogeneous along the direction of the incident radiation (Fig. S3). A more intricate calculation would require knowledge on the particle size and relative particle size distribution \cite{Kayser2015}, as well as on the surface coverage \cite{Unterumsberger2020}. Under TXRF conditions, the mass deposition $m_{\text{A},k}$ of element $k$ for each incidence angle $\theta$ can then be determined from the respective measured XRF count rate \cite{Beckhoff2007}
\begin{equation}
    m_{\text{A},k}=\frac{-1}{\mu_{eff}(E_0,E_k)}\ln\left({1-\frac{I_k (\theta)\sin\theta\,\mu_{eff}(E_0,E_k)}{\frac{\Omega}{4\pi}\, I_0 \,\overline{XSW} (\theta)\,\omega_k\,\tau_k (E_0 )\,\epsilon(E_k)}}\right)
    \label{eq:txrf}
\end{equation}
where $\omega_k$ corresponds to the fluorescence yield and $\tau_k (E_0 )$ to the photoionization cross-section of the element (index $k$) being quantified. The values of atomic fundamental parameters can be found in literature databases \cite{SCHOONJANS2011} or selected parameters are determined in dedicated experiments as for the fluorescence yield for C \cite{Beckhoff2001} or O \cite{Honicke2016}. The factor $\mu_{eff} (E_0,E_k)$ accounts for the effective absorption cross-section of incident and emitted X-ray photons labelled $\mu_{in} (E_0 )$ and $\mu_{out} (E_k )$ respectively, within the PM investigated
\begin{equation}
 \mu_{eff}(E_0,E_k)=\sum_j c_j \left(\frac{\mu_{in,j} (E_0 )}{\sin\theta}+\frac{\mu_{out,j} (E_k )}{\sin \frac{\pi}{2}-\theta}\right)   
\end{equation}
and requires hence knowledge on the mass deposition of the different elements present in order to take correctly into account the relative contributions via the factor $c_k=\frac{\overline{m_{\text{A},k}}}{\sum_j \overline{m_{\text{A},j}}} $ with $\overline{m_{\text{A},k}}$ being the mean quantified mass deposition at incidence angles above the critical angle for total external reflection (more precisely from 6$^\circ$ to 10$^\circ$) where no XSW is present ($\overline{XSW} (\theta) =1$ for $\theta > 3\, \theta_c$). In this angular regime it can also be shown by means of a first order Taylor expansion of Eq. \ref{eq:txrf} that the quantification result will correspond to the one from a standard XRF quantification as used in Ref. \cite{Beckhoff2008}.

A consideration which is usually made at larger incidence angles during the quantification is the correction for absorption of X-rays on the incidence and emission paths 
\begin{equation}
    M_k(E_0,E_k)=\frac{\sum_j \overline{m_{\text{A},j}} (\frac{\mu_{in,j} (E_0 )}{\sin \theta}+\frac{\mu_{out,j} (E_k )}{\sin \frac{\pi}{2}-\theta})}{1-\exp({-\sum_j \overline{m_{\text{A},j}} (\frac{\mu_{in,j} (E_0 )}{\sin \theta}+\frac{\mu_{out,j} (E_k )}{\sin \frac{\pi}{2}-\theta})})}
\end{equation}
It was found that for incidence angles in the range from 6$\circ$ to 10$\circ$ this factor accounts for at most a few percent only (less than 5$\%$) for most of the samples. Only for samples with very high mass depositions a relative correction of 25$\%$ to 30$\%$ was introduced in this iterative correction scheme. For a most accurate correction factor and quantification a complete knowledge of the matrix composition is required. Furthermore, secondary fluorescence due to photoelectrons or fluorescence is neglected. This introduces only a minor error for low mass depositions but, depending on the matrix composition and the incident photon energy, should not be disregarded for high mass depositions where errors of up to 20$\%$-30$\%$ can be introduced \cite{Wahlisch2020}. 

Finally, the GIXRF measurement allows to compare the quantification results of the elemental mass deposition $m_{\text{A},k}$ between TXRF conditions and XRF conditions under shallow incidence angles. The uncertainty made in the quantification depends on the uncertainties on the incident flux ($1\%$), the XSW factor ($5\%$ in the angular range where it needs to be taken into account), the atomic fundamental parameters (fluorescence yield, $10\%$ for light elements, and photoionization cross-section, $7.5\%$), the detector efficiency and spectral deconvolution ($2.5\%$), the counting statistics and the solid angle of detection (about $15\%$ for the smallest incidence angles to about $4\%$ for the largest incidence angles used for quantification)\cite{Beckhoff2007}. Thus, the systematic errors, which disregard any sample effects, usually amount to about $20\%$ in in the TXRF regime and to about $12\%$ or better for the largest incidence angles used in the GIXRF measurements.

Note, that the mass deposition in terms of mass (or likewise number of atoms) for each element per unit area is quantified. A conversion to mass, which is a more commonly used metric in the aerosol community, can be straightforwardly realized if the area on which the airborne PM is collected and its lateral distribution are known.

\section{Results \& Discussion}
Given the uniform distribution of the PM, quantification results by means of Eq. \ref{eq:txrf} can be expected to be constant for each incidence angle covered throughout a GIXRF measurement. Indeed, the different excitation conditions realized at each incidence angle should only affect the lowest limit of detection achievable due to a changing and even vanishing XSW and an increasing penetration depth into the bulk volume of the substrate but not impact the quantification result. 

\begin{figure}
	\begin{center}
	\includegraphics[width=0.8\textwidth]{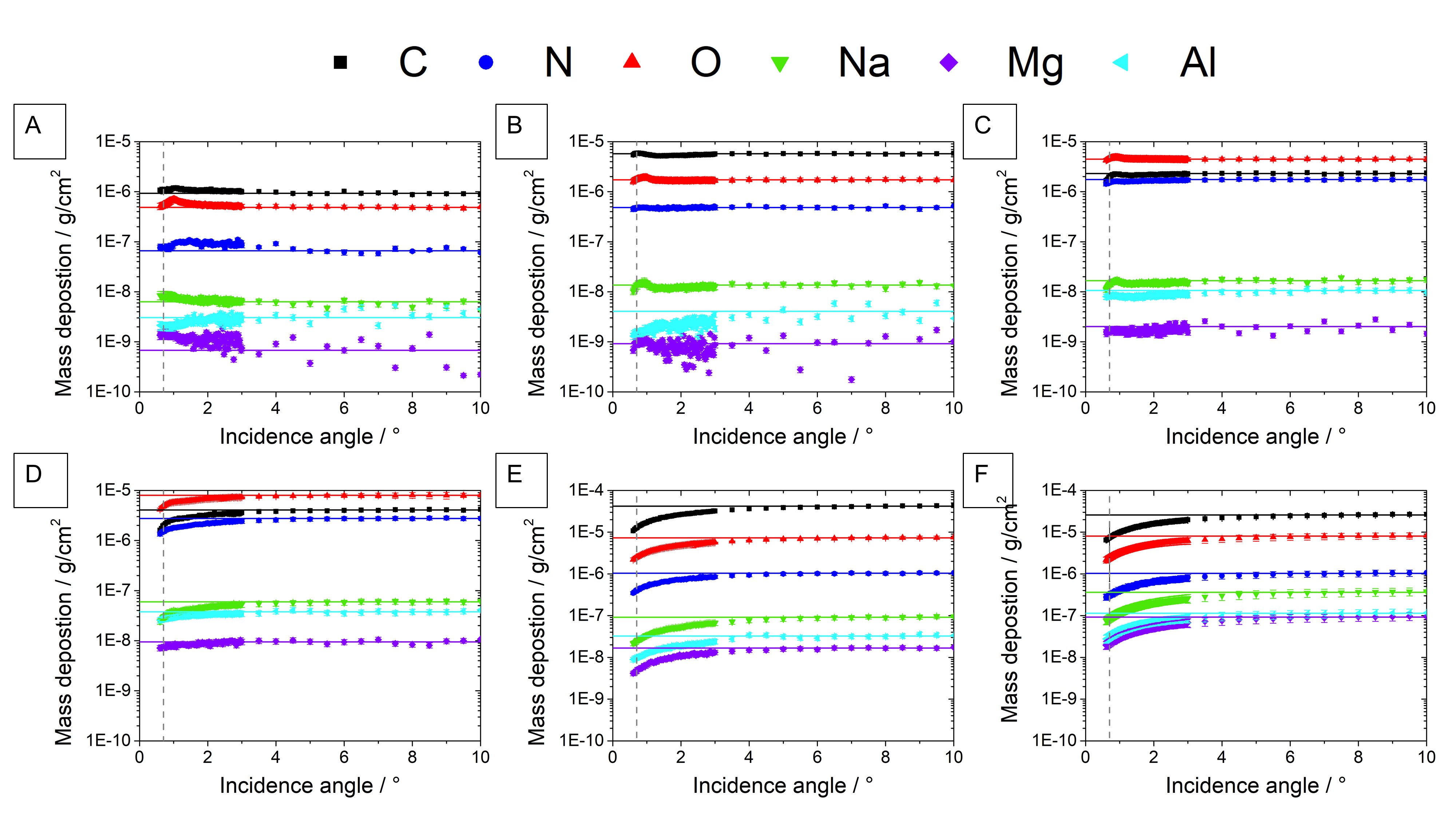}
	\end{center} 
	\caption{Quantified mass deposition for the different incidence angles covered when varying the excitation conditions during the GIXRF measurement from the TXRF regime to the XRF regime under shallow incidence angles. The vertical bar indicates the position typically selected for a TXRF measurement for a Si substrate and the incident photon energy used, while the horizontal bar indicate the mass deposition quantified at the largest incidence angles for each element. For low mass depositions (upper 3 panels) a satisfyingly good agreement can be observed, but for increasing mass deposition a growing discrepancy appears for all the elements. This indicates that not all physical effects due to attenuation of X-rays in the collected airborne PM are accounted for in the quantification scheme. Under shallow incidence angles attenuation is less important and has therefore a lesser impact on the quantification scheme as can be seen from the results approaching a constant value.}
	\label{fig:figure4}
\end{figure}

However, a consistent quantification result where comparable elemental mass depositions are quantified for all incidence angles used is not achieved for all samples. More explicitly, only for a part of the samples present a good agreement under TXRF conditions, $\theta \approx 0.7^\circ$, and under XRF conditions, $\theta > 3.0^\circ$  (Fig. \ref{fig:figure4}). For the lowest mass depositions used, the quantification results are independent of the incidence angle and agree reasonably well with each other (Fig. \ref{fig:figure4}, upper panels). In this situation, the quantification of the mass deposition does not depend on the excitation conditions used such that the GIXRF measurement is useful to validate the results from the TXRF measurement, which is usually performed at a single position. This observation is congruent with the proven reliability of TXRF for quantifying trace level contamination, as it is routinely done for semiconductor applications. Nonetheless, the GIXRF data allows providing more robust quantification data as it allows to inherently validate results obtained by means of TXRF. A specificity can be observed for samples A and B where an imperfect deconvolution of the XRF spectra recorded at larger incidence angles affects the quantification results because of the underlying Si-RRS which is not perfectly described by the model used. In particular for Al, whose main characteristic line is close to the high energy cut-off of the Si RRS at 1520 eV, but partially also for Mg this results as well in a larger scattering of the quantification results at larger incidence angles (Fig. \ref{fig:figure1}). This issue illustrates perfectly the main benefit of TXRF for low mass depositions since it allows suppressing background contributions from the substrate (Fig. \ref{fig:figure1}).

For higher mass loadings, a discrepancy between the quantification results appears (Fig. \ref{fig:figure4}, lower panels) in the sense that under TXRF conditions mass depositions are underestimated. For these samples, the quantification results are not consistent throughout the monitored angular range such that the GIXRF measurement falsifies the TXRF results. These samples indicate that it is necessary to be aware of the range of validity of TXRF quantification results and to validate findings on unknown samples by a GIXRF measurement where the excitation conditions are varied in a controlled manner by varying the incidence angle. While the comparison of quantified mass depositions on each sample allows assessing whether the robustness of the quantification results obtained under TXRF conditions, GIXRF allows extending the dynamic range of mass depositions which can be quantified. Indeed, the variation from shallow to larger incidence angles still provides a considerable variation of the path length of the incident photons through the airborne PM such that absorption effects, which may affect the validity of quantification results, are probed while the premise of an XSW which is not or only to a little extent perturbed by the presence of airborne PM on the substrate is not given. The influence of the airborne PM on the XSW is also qualitatively visible in the GIXRF measurement: for the samples with the lowest mass deposition an enhanced XRF rate was detected in the vicinity of the critical angle of total external reflection compared to larger incidence angles, while for the highest mass the opposite was the case. Hence, in the latter case no XSW was created and by this way the GIXRF measurement by itself indicates already that quantification under TXRF conditions is compromised despite the fact that the measured reflectivity is used in the quantification (Fig. \ref{fig:figure5}).

If it was not taken into account how increasing mass deposition affect the contrast in optical density at the interface defined by the surface of the substrate the discrepancy between quantification results for the different incidence angles used would even be more important (Eq. \ref{eq:txrf}). This insight emphasizes the benefit of monitoring in parallel to a TXRF or GIXRF measurement the reflectivity from the sample. The dependence of the XSW on the surface coverage must be taken into account when quantifying the mass deposition, a statement which is not only valid when using the reference-free quantification approach but also when applying external standards. For the extreme case of the two samples with the highest PM loads (panels E and F in Figs. \ref{fig:figure2} and \ref{fig:figure4}) the variation of the quantified mass deposition with the incident angle indicates that an accurate quantification is tedious since here the attenuation of the incident radiation within the PM collected would need to be considered. This aspect introduces considerable uncertainties in the final result. Hence, a GIXRF measurement allows to discard these types of samples from further use in analytical campaigns.

The relationship between the mass deposition and the XSW is also noteworthy (Eq. \ref{eq:txrf}) with regard to the need for using representative specimen when applying external standards for quantification purposes. In case of internal standards, reliable results are only obtained if the collected mass deposition of the airborne PM is within the range of mass depositions covered by the standard under the premise that a homogeneous intermixing is realized. In other words, the dynamic range within which the calibration is valid needs to be considered. Finally, in the reference-based approaches the XSW needs to be comparable between the calibration material and investigated sample material, be it locally when using an internal standard or between samples when using an external standard, in order to avoid a calibration bias. An upper limit for reliable TXRF quantification is discussed in literature in terms of critical thickness \cite{KLOCKENKAMPER1989} and saturation effect \cite{Hellin2004}.

A further reason for the deviation of TXRF quantification results as compared to the results obtained at the largest incidence angles in the GIXRF scan is that the full volume of the PM collected is not illuminated homogeneously in its depth direction because of the attenuation of the incident and reflected X-ray radiation. For increasing incidence angles and high surface coverage the effective path length is reduced as $\frac{1}{\sin\theta}$ such that the X-ray attenuation within the PM volume becomes less pronounced. For samples with a high surface density of airborne PM, this argument becomes even more crucial under conditions where an XSW is expected since then the effective path lengths of the incident and reflected X-rays need to be considered. This insight impacts directly the reliability of quantification by means of TXRF and it becomes necessary to indicate an upper limit for the range of validity of TXRF quantification results. A challenge in air quality monitoring campaigns is that not all stages from a sampling interval will be affected the same as important variations between the different stages can be expected due to inhomogeneous particle size distribution of airborne PM. In case all stages are affected such a situation is the result of exceedingly long collection times or high pollution levels. 

\begin{figure}[tbp]
  \centering
  \begin{minipage}{1.0\linewidth}
    \centering
    \includegraphics[width=0.45\textwidth]{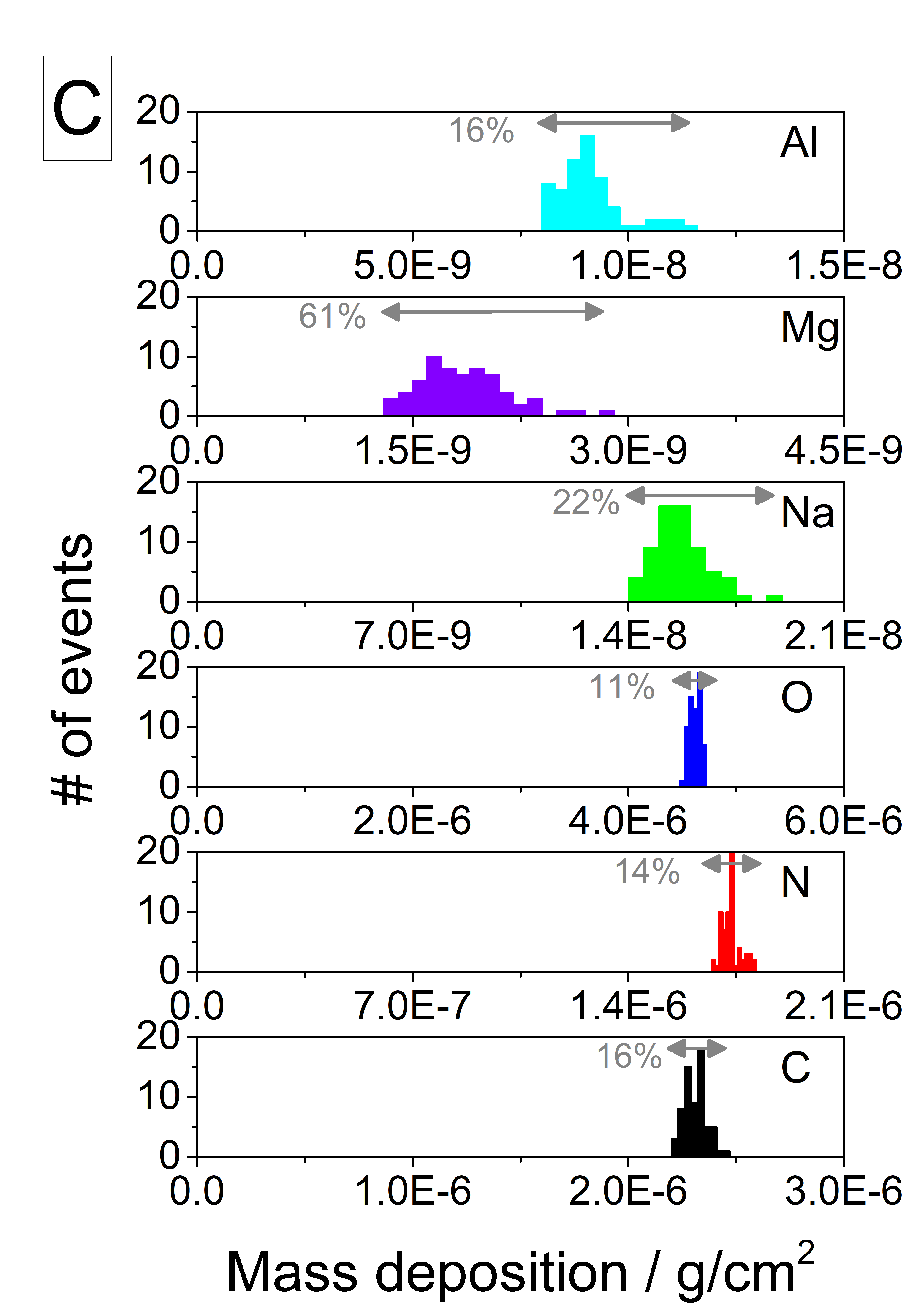}
    \quad
		\includegraphics[width=0.45\textwidth]{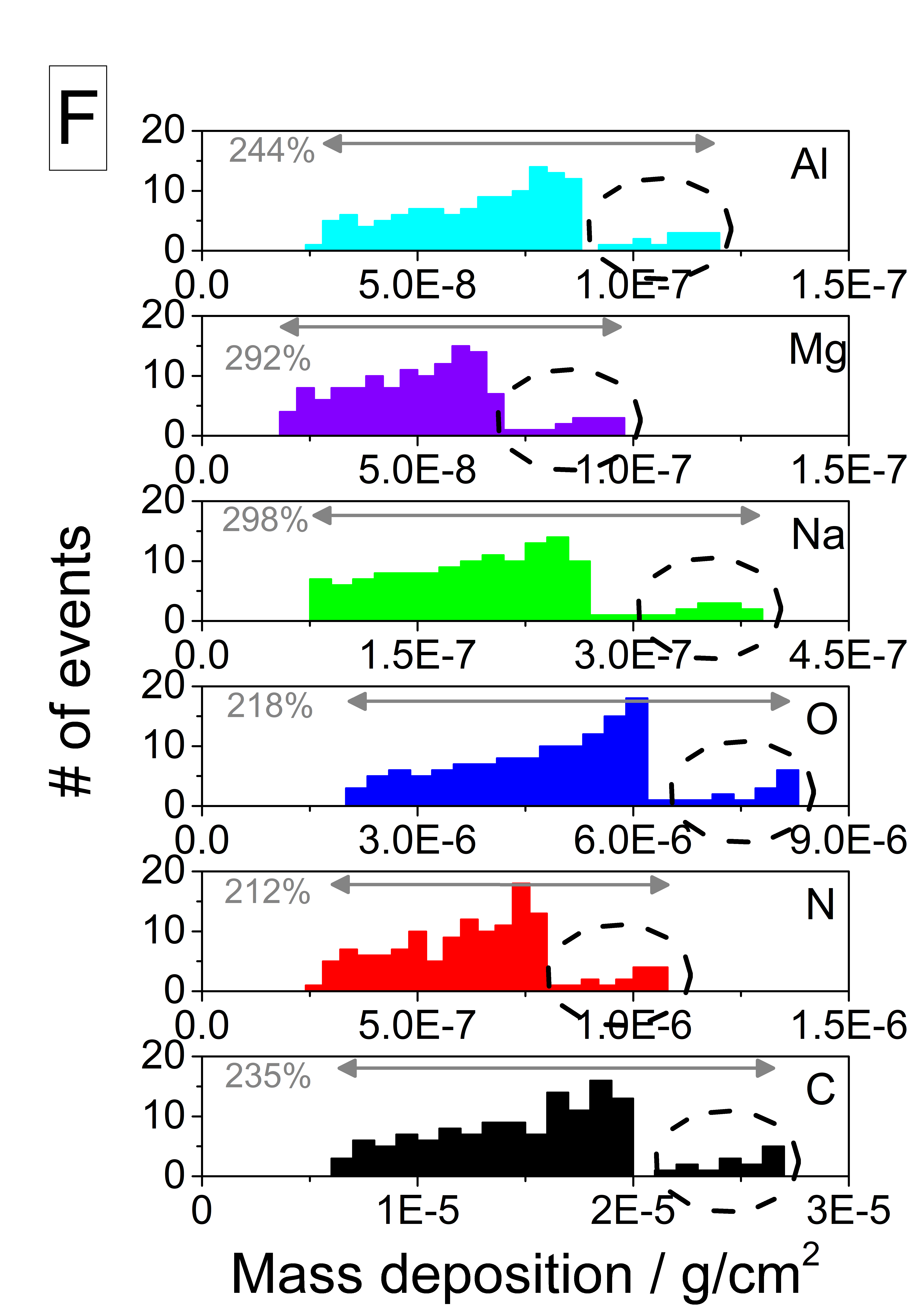}
  \end{minipage}
  \vspace{.5ex}
  \par
  \begin{minipage}{1.0\linewidth}
	\renewcommand{\tabcolsep}{1pt}
		\centering
		\begin{tabular}{|c|c|c|c|c|}
			\toprule
      {\tiny Element} & {\tiny $m_{\text{TXRF}}$ / $\frac{\text{g}}{\text{cm}^2}$} & {\tiny $\sigma_{\text{TXRF}}$ / $\frac{\text{g}}{\text{cm}^2}$}  & {\tiny $m_{\text{GIXRF}}$ / $\frac{\text{g}}{\text{cm}^2}$} &  {\tiny $\sigma_{\text{GIXRF}}$ / $\frac{\text{g}}{\text{cm}^2}$}\\
			\hline
      C & 0.7 E-5 & 0.5 E-5 & 2.6 E-5 & 1.0 E-5 \\
			N & 0.3 E-6 & 0.1 E-6 & 1.0 E-6 & 0.4 E-6 \\
			O & 2.4 E-6 & 1.0 E-6 & 8.0 E-6 & 3.2 E-6 \\
			Na & 0.9 E-7 & 0.5 E-7 & 3.6 E-7 & 1.9 E-7 \\
			Mg & 2.3 E-8 & 1.2 E-9 & 9.2 E-8 & 4.8 E-8 \\
			Al & 0.3 E-7 & 0.2 E-7 & 1.1 E-7 & 0.6 E-7 \\
      \bottomrule
    \end{tabular}
		\quad
		\begin{tabular}{|c|c|c|c|c|}
		\toprule
     {\tiny Element} & {\tiny $m_{\text{TXRF}}$ / $\frac{\text{g}}{\text{cm}^2}$} & {\tiny $\sigma_{\text{TXRF}}$ / $\frac{\text{g}}{\text{cm}^2}$}  & {\tiny $m_{\text{GIXRF}}$ / $\frac{\text{g}}{\text{cm}^2}$} &  {\tiny $\sigma_{\text{GIXRF}}$ / $\frac{\text{g}}{\text{cm}^2}$}\\
			\hline
      C & 1.9 E-6 & 0.2 E-6 & 2.3 E-6 & 0.2 E-6 \\
			N & 1.5 E-6 & 0.1 E-6 & 1.8 E-6 & 0.2 E-6 \\
			O & 4.6 E-6 & 0.4 E-6 & 4.5 E-6 & 0.4 E-6 \\
			Na & 1.4 E-8 & 0.2 E-8 & 1.7 E-8 & 0.2 E-8 \\
			Mg & 1.6 E-9 & 0.2 E-9 & 2.0 E-9 & 0.3 E-9 \\
			Al & 0.8 E-8 & 0.1 E-8 & 1.1 E-8 & 0.2 E-8 \\
      \bottomrule
    \end{tabular}
  \end{minipage}
  \caption{Histogram of the GIXRF quantification results for samples C and F. Indicated above the vertical bars is the difference between the $5\%$ and $95\%$ percentile relative to the $5\%$ value. For sample C an agreement throughout the full angular range, i.e., with and without XSW, is obtained, while sample F TXRF underestimates the elemental mass depositions. This becomes evident by a comparison to the quantification results obtained under the XRF regime under shallow incidence angles (dashed circles), where no XSW needs to be considered ($\theta > 3\,\theta_c$). The tables include the quantification results under TXRF conditions and XRF conditions using shallow incidence angles. The $\sigma$ value does not represent the systematic quantification error, but the standard deviation of the results obtained in the TXRF regime (incidence angles from 0.6$^\circ$ to 0.8$^\circ$) and XRF conditions using shallow incidence angles (incidence angles from 6$^\circ$ to 10$^\circ$). A ratio of up to a factor 4 between the different results can be observed.}
  \label{fig:figure6}
\end{figure}

Depending on the amount of collected airborne PM, the agreement respectively discrepancy in the quantification results obtained for the different excitation conditions achieved in a GIXRF measurement becomes even more obvious when considering the distribution of the quantified mass deposition for each element of a sample (Fig. \ref{fig:figure6}). The results for the further samples are included in Figs. S4 and S5.  In Fig. \ref{fig:figure6} the relative range between the lower and upper 5$^\text{th}$ percentile are indicated. For sample C the results for each element show a very good consistency with each other, mostly within a range of several percent only as can also be recognized from the tabulated values. This agreement can be considered as acceptable for air quality monitoring campaigns. For the sample with the highest mass depositions (sample F), the relative difference between the lower and upper 5$^\text{th}$ percentile relative to the mean value amounts to about a factor of 3 to 4 depending on the sample considered. Given that the measurement were not equidistant in the incidence angles, the results at the lower incidence angles have a higher impact on the percentile intervals and the discrepancy between the quantification results can even be larger than indicated. This aspect is highlighted by the dashed circles in Fig. \ref{fig:figure6} which indicate the quantification results obtained at the largest incidence angles which clearly indicate the importance of the deviation obtained from the quantification performed in the TXRF regime. This discrepancy makes the need for self-consistent validation by means of GIXRF measurements evident: the more exhaustive data obtained by quantifying the mass deposition for different incidence angles and excitation conditions reveals immediately any discrepancies in the quantification under TXRF conditions when for the sample considered deviations in the quantification results are observed. In view of the demands on quantitative techniques for regulatory purposes this critical assessment of the validity of the results is mandatory.  

As a consequence, we propose to expand TXRF based quantitative analysis of airborne PM to GIXRF. From an experimental point of view this approach allows combining the benefit of TXRF (low limits of detection) with the ones of GIXRF (higher dynamic range of mass depositions that can be covered, reliable quantification for higher mass depositions) since total reflection excitation conditions are necessarily covered during a GIXRF measurement. By this means, unknown amounts of airborne PM can optimally be handled since low mass depositions can be quantified in the angular regime below and high deposition samples can in the angular regime above the critical angle of total external reflection. In both cases the GIXRF measurements allow for robust and reliable quantification with internal validation by assessing whether the quantification results are in reasonable agreement. It can be noted that for this purpose a sparser set of incidence angles can be monitored in order to arrange for more time-efficient measurements as required for example during field campaigns where high-throughput quantification of airborne PM is aimed at. From an instrumental point of view this strategy is nowadays a solution which can be readily implemented, as more and more successful examples for laboratory GIXRF instruments and even commercially available GIXRF instruments exist. In case only TXRF measurements can be performed, the consequence is that only a restricted range of samples or mass depositions can be investigated. Indeed, for reliable quantification under TXRF conditions the attenuation of X-rays within the collected airborne PM and the differences in the XSW created should remain below a threshold value. This criteria cannot be specified generically since it depends on the incident photon energy and the absolute elemental composition of the PM.

Furthermore, simple control cross-check measurements can be used to identify possible issues with the quantification performed, regardless if standards are used or a reference-free quantification scheme is being applied. One indicator is the reflectivity from the substrate as compared to a blank substrate of the same type. This cross-check can even be applied when samples are investigated under TXRF conditions solely (Fig. \ref{fig:figure4}). An other possibility is to consider the angular intensity profile of the XRF originating from the bulk volume of the substrate and to compare it against the one of a blank substrate (Fig. S6). For increasing mass deposition of airborne PM and larger incidence angles significant attenuation compared to a blank Si substrate can be observed. In both cases the comparison to a blank substrate allows elucidating whether, besides possible sample alignment issues, quantification by TXRF alone is compromised due to the amount of airborne PM collected exceeding the range within which TXRF can be validly applied. 

\section{Conclusion \& Outlook}
It was shown that GIXRF by means of a controlled variation of the excitation conditions allows for applying a robust quantification scheme and, hence, for assessing in a self-consistent manner the validity of quantification under TXRF conditions over a wide range of mass depositions. Self-consistent means that throughout the different excitation conditions covered during a GIXRF measurement the quantification results are in agreement with each other for each incidence angle in case of low mass depositions and for incidence angles far above the critical angle of total reflection for higher mass depositions. Varying pollution levels and particle size repartition make such verification of the validity of results necessary since reliable quantification in the TXRF regime can only be realized for mass depositions on the level of lowest amount of matter where a linear calibration between count rate and mass deposition can be guaranteed. This range depends on the matrix composition, the airborne PM size and the incident photon energy, but cannot be assessed from TXRF measurements alone.  While increasing mass depositions result in nonlinear effects in terms of different XSW created on the top of the substrates and pronounced attenuation of the incident and reflected X-rays within the airborne PM collected, such that quantification by means of TXRF alone is compromised. Therefore criteria or thresholds for reliable quantification by means of TXRF need to be established in the future. This pitfall cannot be readily circumvented, let alone be identified by means of calibration samples or procedures since, both, the range of elemental mass deposition and the matrix composition on the substrates are unknown beforehand of any measurement in the field. As a consequence a robust quantification scheme requires GIXRF measurements, i.e., measurements covering two fundamentally different excitation regimes. Only then the quantified elemental mass depositions can be considered as reliable and validated, respectively the range of validity for quantification results obtained by means of TXRF can be assessed for a selected class of samples and experimental conditions. Indeed, it has to be noted that the presented experiment with its emphasis on light elements was realized in the soft X-ray regime but the conclusions made can also be applied for higher X-ray energies. 

While advanced instrumentation was applied for allowing for a physically traceable quantification which does not rely on the use of standards, it can be emphasized that this approach is transferable to laboratory instrumentation \cite{HONICKE2020}. Thus, possibilities are offered to transfer the approach presented to instrumentation used for high-throughput measurements in the field or in the laboratory. The additional implementation of a diode to measure the reflectivity would not only allow for more accurate determination of the XSW but would also provide a rough but straightforward indication whether the quantity of airborne PM collected presents an issue for the quantification by means of TXRF.

\section*{Acknowledgments}
Parts of this research was performed within the EMPIR project 19ENV08 AEROMET II. This project has received funding from the EMPIR programme co-financed by the Participating States and from the European Union’s Horizon 2020 research and innovation programme. The support of the European Structural and Investment Funds jointly financed by the European Commission and the Hungarian Government through grant no. VEKOP-2.3.2-16-2016-00011 (on behalf of J. Os\'{a}n) is also appreciated.

\cleardoublepage
\bibliography{biblio}

\cleardoublepage
\include{ESI.pdf}

\end{document}


\maketitle

\newpage

\begin{figure}
	\begin{center}
	\includegraphics[width=0.8\textwidth]{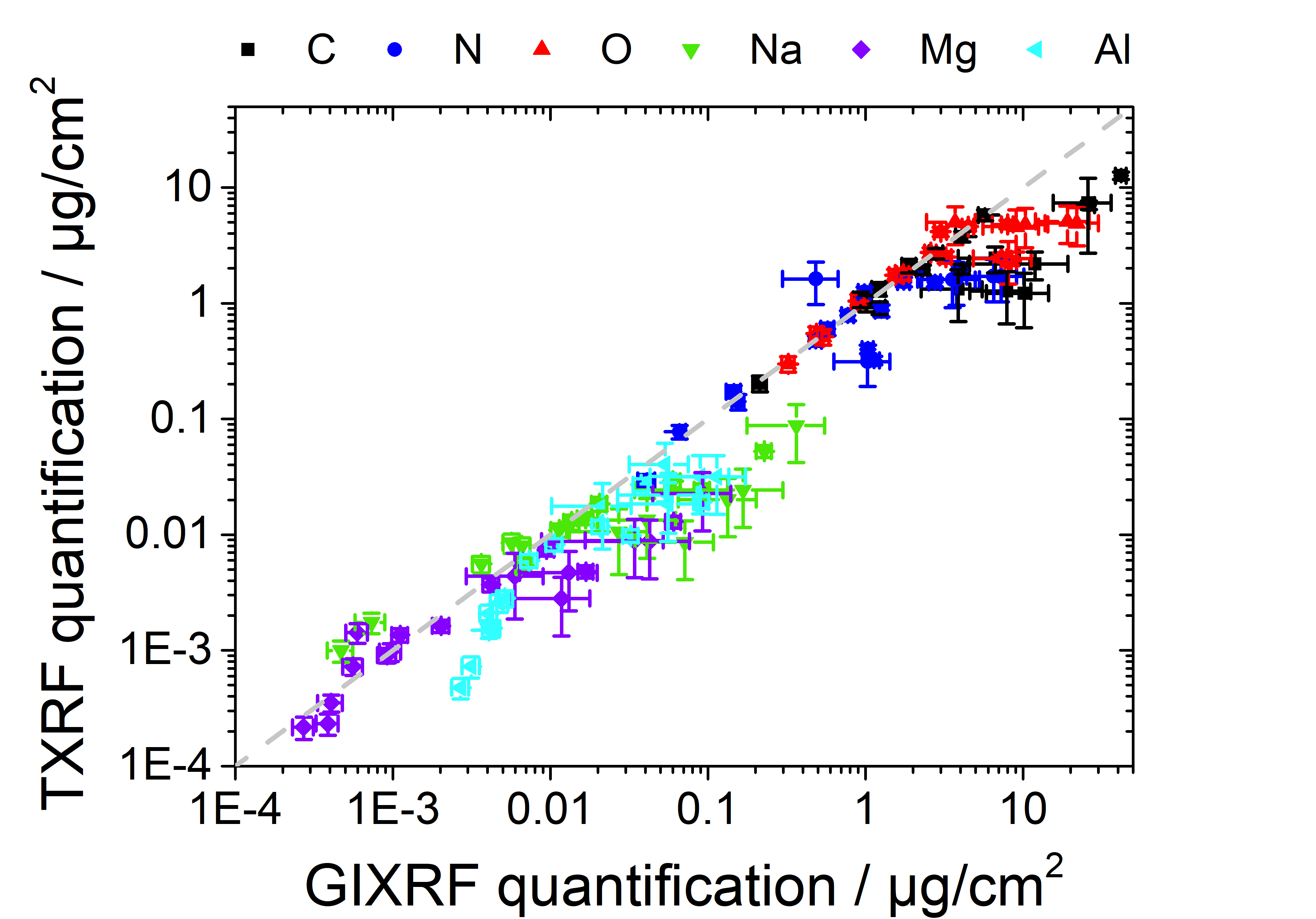}
	\end{center} 
	\caption{Comparison for all 19 samples investigated of the quantified mass deposition for TXRF and XRF (under shallow incidence angles, no XSW is created) excitation conditions. The different excitation conditions are covered during the GIXRF measurement. The deviation from the expected ratio with increasing elemental mass deposition is evident for all elements. The horizontal bar indicates the average mass deposition quantified at the largest incidence angles used and the vertical bar the incidence angle at which a TXRF quantification is typically performed for Si wafers and the incident photon energy used. In the case of low mass depositions, the expected 1:1 ratio indicated by the dashed line is matched but for increasing mass deposition a deviation is observed which indicates that under TXRF conditions the quantified mass deposition underestimates the actually deposited mass deposition. The error bars include the quantification error and the standard deviation of quantification results in the respective angular range considered. Such a deviation from a consistent quantification of the mass deposition does not become apparent from a TXRF measurement itself, as it is a single-point measurement, and more exhaustive data as collected during a GIXRF measurement is required. A better representation would be using the total mass, but due to the soft X-ray radiation used it can not be ascertained that the XRF radiation for all elements present in the matrix was excited.}
	\label{fig:figure6}
\end{figure}

\newpage
 
\begin{figure}
	\begin{center}
	\includegraphics[width=0.8\textwidth]{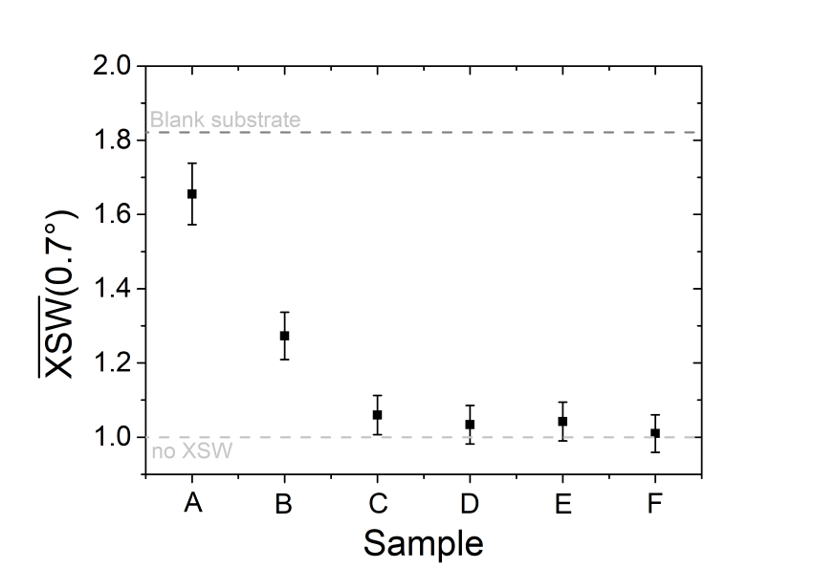}
	\end{center} 
	\caption{An increasing mass deposition affects the contrast in optical density at the interface defined by the surface of the substrate: while for a low mass deposition the surface coverage is low enough to assume that the airborne PM does not alter the optical properties in the area above the substrate surface, this assumption does not hold for high mass deposition and the optical properties of the airborne PM for the incident X-ray radiation needs to be considered in addition to optical properties of the substrate. based on the measured reflectivity at an incidence angle of 0.7$^\circ$ (as shown in Fig. 4, vertical bar), significant differences in the $\overline{XSW}(\theta)$ between the different samples are obtained. The horizontal bars indicate the relative XSW factor which can be expected from the reflectivity measured from a blank Si substrate without PM on the top of it, respectively in the case no XSW is present because the reflectivity has dropped to 0. Finally, the measured reflectivity can be used as a monitor for assessing whether comparable excitation conditions are achieved for a sample set under investigation.}
	\label{fig:figure2}
\end{figure}

\newpage

\begin{figure}
	\begin{center}
		\includegraphics[width=0.8\textwidth]{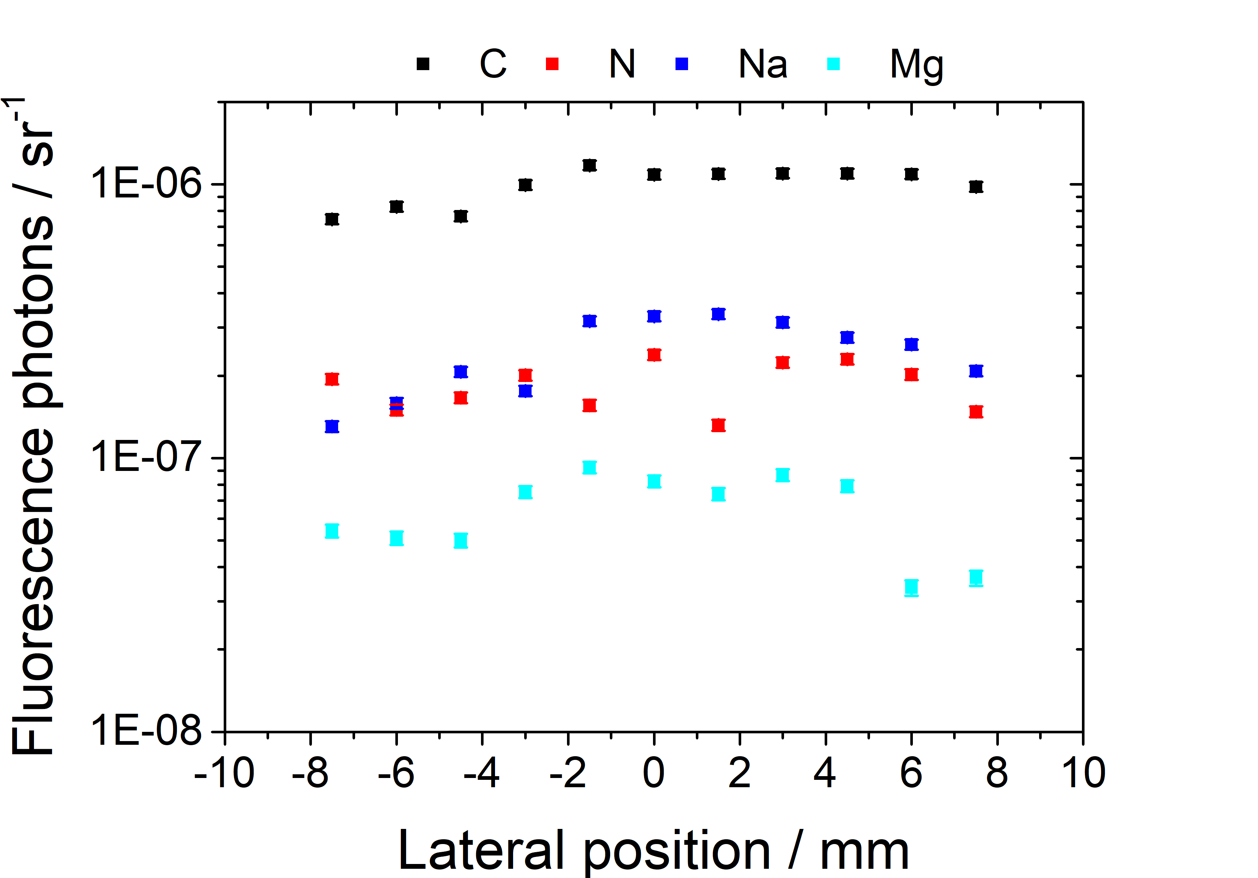}
	\end{center} 
	\caption{Lateral scan along the line of the deposition pattern exhibiting that in the central area a quite uniform distribution has been achieved. Due to the shallow incidence angle of 5$^\circ$, the position close to the boarder of the sample could not be investigated but no decreased collection efficiency is expected from design consideration of the May-type cascade impactor. The sample is the one for which the GIXRF data is displayed in Fig. 1. }
	\label{fig:figure3}
\end{figure}
 
\newpage

\begin{figure}
	\begin{center}
		\includegraphics[width=\textwidth]{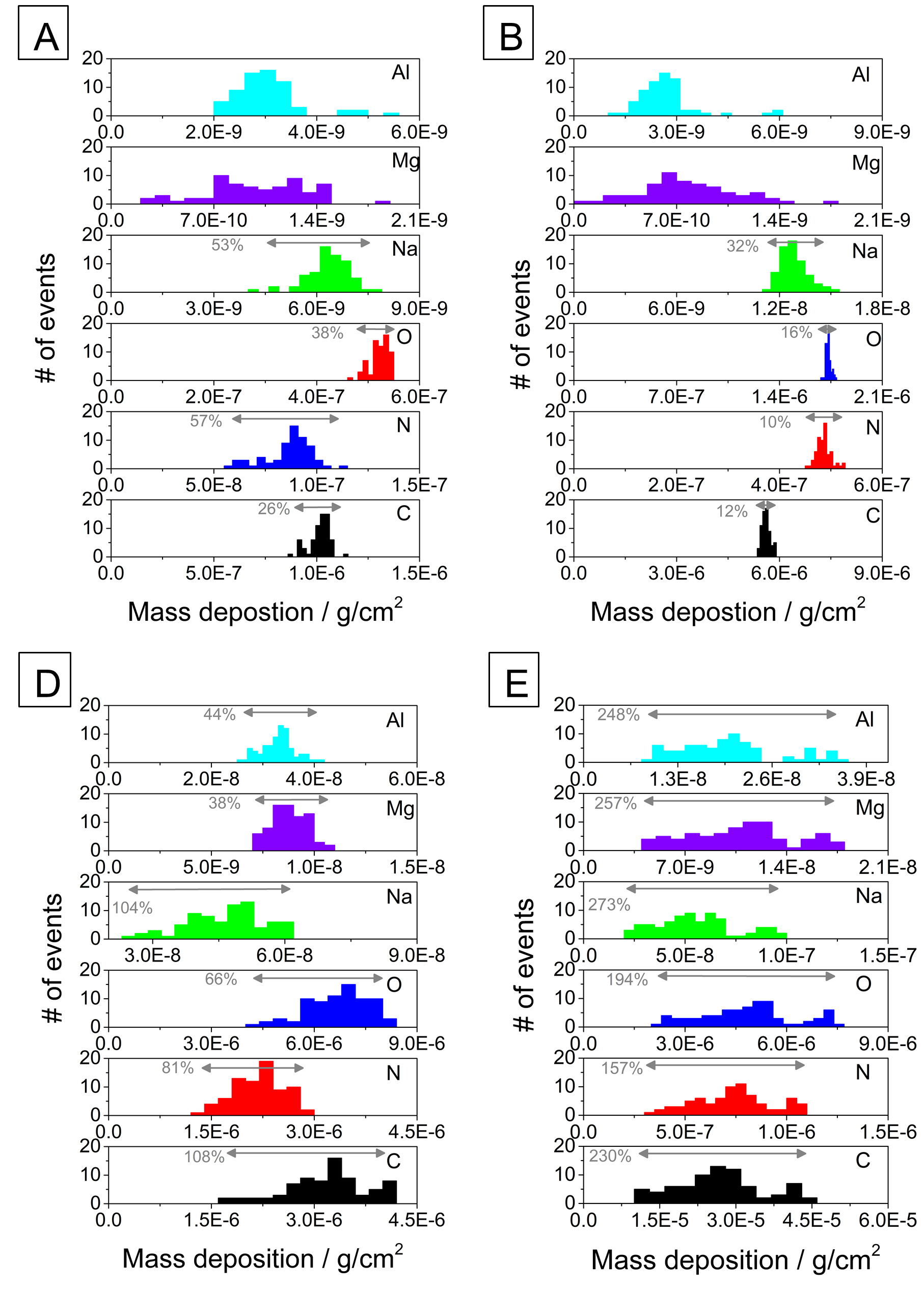}
	\end{center} 
	\caption{Distribution of the quantified mass depositions of the different elements for samples A, B, D and E.  Indicated above the vertical bars is the difference between the $5\%$ and $95\%$ percentile relative to the $5\%$ value. This illustrates the agreement, respectively growing discrepancy between quantification results obtained under different excitation conditions with increasing mass deposition.}
	\label{fig:figure7x}
\end{figure}

\newpage

\begin{figure}[tbp]
  \centering
  \begin{minipage}{1.0\linewidth}
    \centering
    \begin{tabular}{|c|c|c|c|c|}
      \toprule
       Element &  $m_{\text{TXRF}}$ / $\frac{\text{g}}{\text{cm}^2}$ & $\sigma_{\text{TXRF}}$ / $\frac{\text{g}}{\text{cm}^2}$ & $m_{\text{GIXRF}}$ / $\frac{\text{g}}{\text{cm}^2}$ &  $\sigma_{\text{GIXRF}}$ / $\frac{\text{g}}{\text{cm}^2}$ \\
			\hline
      C & 1.1 E-6 & 0.2 E-6 & 0.9 E-6 & 0.1 E-6 \\
			N & 7.8 E-8 & 1.0 E-8 & 6.6 E-8 & 0.6 E-8  \\
			O & 5.5 E-7 & 0.8 E-7 & 4.9 E-7 & 0.4 E-7 \\
			Na & 8.5 E-9 & 1.6 E-9 & 5.7 E-9 & 0.7 E-9 \\
			Mg & 1.4 E-9 & 0.3 E-9 & \textit{0.6 E-9} & \textit{1.0 E-9} \\
			Al & 2.1 E-9 & 0.4 E-9 & \textit{4.7 E-9} & \textit{0.6 E-9} \\
      \bottomrule
    \end{tabular}		
  \end{minipage}
  \vspace{.5ex}
  \par
  \begin{minipage}{1.0\linewidth}
    \centering
    \begin{tabular}{|c|c|c|c|c|}
      \toprule
       Element &  $m_{\text{TXRF}}$ / $\frac{\text{g}}{\text{cm}^2}$ & $\sigma_{\text{TXRF}}$ / $\frac{\text{g}}{\text{cm}^2}$ & $m_{\text{GIXRF}}$ / $\frac{\text{g}}{\text{cm}^2}$ &  $\sigma_{\text{GIXRF}}$ / $\frac{\text{g}}{\text{cm}^2}$ \\
			\hline
      C & 5.8 E-6 & 0.6 E-6 & 5.7 E-6 & 0.5 E-6 \\
			N & 4.7 E-7 & 0.5 E-7 & 4.8 E-7 & 0.4 E-7 \\
			O & 1.8 E-6 & 0.2 E-6 & 1.7 E-6 & 0.2 E-6 \\
			Na & 1.3 E-8 & 0.2 E-8 & 1.4 E-8 & 0.2 E-8 \\
			Mg & 9.1 E-10 & 1.4 E-10 & 9.2 E-10 & 1.2 E-10 \\
			Al & 1.5 E-9 & 0.2 E-9 & \textit{4.1 E-9} & \textit{0.5 E-9} \\
      \bottomrule
    \end{tabular}
  \end{minipage}
	\vspace{.5ex}
  \par
  \begin{minipage}{1.0\linewidth}
    \centering
    \begin{tabular}{|c|c|c|c|c|}
      \toprule
       Element &  $m_{\text{TXRF}}$ / $\frac{\text{g}}{\text{cm}^2}$ & $\sigma_{\text{TXRF}}$ / $\frac{\text{g}}{\text{cm}^2}$ & $m_{\text{GIXRF}}$ / $\frac{\text{g}}{\text{cm}^2}$ &  $\sigma_{\text{GIXRF}}$ / $\frac{\text{g}}{\text{cm}^2}$ \\
			\hline
      C & 2.0 E-6 & 0.2 E-6 & 4.1 E-6 & 0.4 E-6 \\
			N & 1.5 E-6 & 0.1 E-6 & 2.7 E-6 & 0.2 E-6 \\
			O & 4.8 E-6 & 0.4 E-6 & 8.0 E-6 & 0.7 E-6 \\
			Na & 2.9 E-8 & 0.3 E-8 & 6.0 E-8 & 0.7 E-8 \\
			Mg & 7.4 E-9 & 0.9 E-9 & 9.5 E-9 & 1.0 E-9 \\
			Al & 2.7 E-8 & 0.3 E-8 & 3.8 E-8 & 0.4 E-8 \\
      \bottomrule
    \end{tabular}
  \end{minipage}
	\vspace{.5ex}
  \par
  \begin{minipage}{1.0\linewidth}
    \centering
    \begin{tabular}{|c|c|c|c|c|}
      \toprule
       Element &  $m_{\text{TXRF}}$ / $\frac{\text{g}}{\text{cm}^2}$ & $\sigma_{\text{TXRF}}$ / $\frac{\text{g}}{\text{cm}^2}$ & $m_{\text{GIXRF}}$ / $\frac{\text{g}}{\text{cm}^2}$ &  $\sigma_{\text{GIXRF}}$ / $\frac{\text{g}}{\text{cm}^2}$ \\
			\hline
      C & 1.3 E-5 & 0.1 E-5 & 4.2 E-5 & 0.3 E-5 \\
			N & 0.4 E-6 & 0.1 E-6 & 1.0 E-6 & 0.1 E-6 \\
			O & 2.5 E-6 & 0.2 E-6 & 7.3 E-6 & 0.6 E-6 \\
			Na & 2.4 E-8 & 0.3 E-8 & 9.2 E-8 & 1.0 E-8 \\
			Mg & 0.5 E-8 & 0.1 E-8 & 1.7 E-8 & 0.2 E-8 \\
			Al & 0.9 E-8 & 0.1 E-8 & 3.2 E-8 & 0.4 E-8 \\
      \bottomrule
    \end{tabular}
  \end{minipage}
  \caption{Quantification results under TXRF conditions and XRF conditions using shallow incidence angles for samples A, B, D and E (top to bottom). The $\sigma$ value does not represent the systematic quantification error, but the standard deviation of the results obtained in the TXRF regime (incidence angles from 0.6$^\circ$ to 0.8$^\circ$) and XRF conditions using shallow incidence angles (incidence angles from 6$^\circ$ to 10$^\circ$). For samples A and B spectral deconvolution was for the larger incidence angles and the elements Na, Mg, Al subject to some error due to Si K resonant Raman scattering. Other than that, the mass depositions quantified in the TXRF regime and in the XRF under shallow incidence angle conditions is in very good agreement. 
	However, the GIXRF measurement reveals for samples D and E a major discrepancy in the results which is due to the fact, due to X-ray attenuation, that under TXRF conditions not all the sample material is probed and the XSW cannot be properly calculated despite measuring the reflectivity from the sample. For XRF under shallow incidence angles, the situation is different (no XSW created, shorter path lengths trough the airborne PM collected) and hence quantification in the TXRF regime is falsified. Note as well that for both excitation regimes, the standard deviation of the results in the angular ranges used is within the systematic errors expected for the quantification. This clearly demonstrates failure of quantification by means of TXRF cannot be uncovered by a TXRF measurement alone.}
  \label{fig:figure6x}
\end{figure}

\newpage

\begin{figure}
	\begin{center}
		\includegraphics[width=0.8\textwidth]{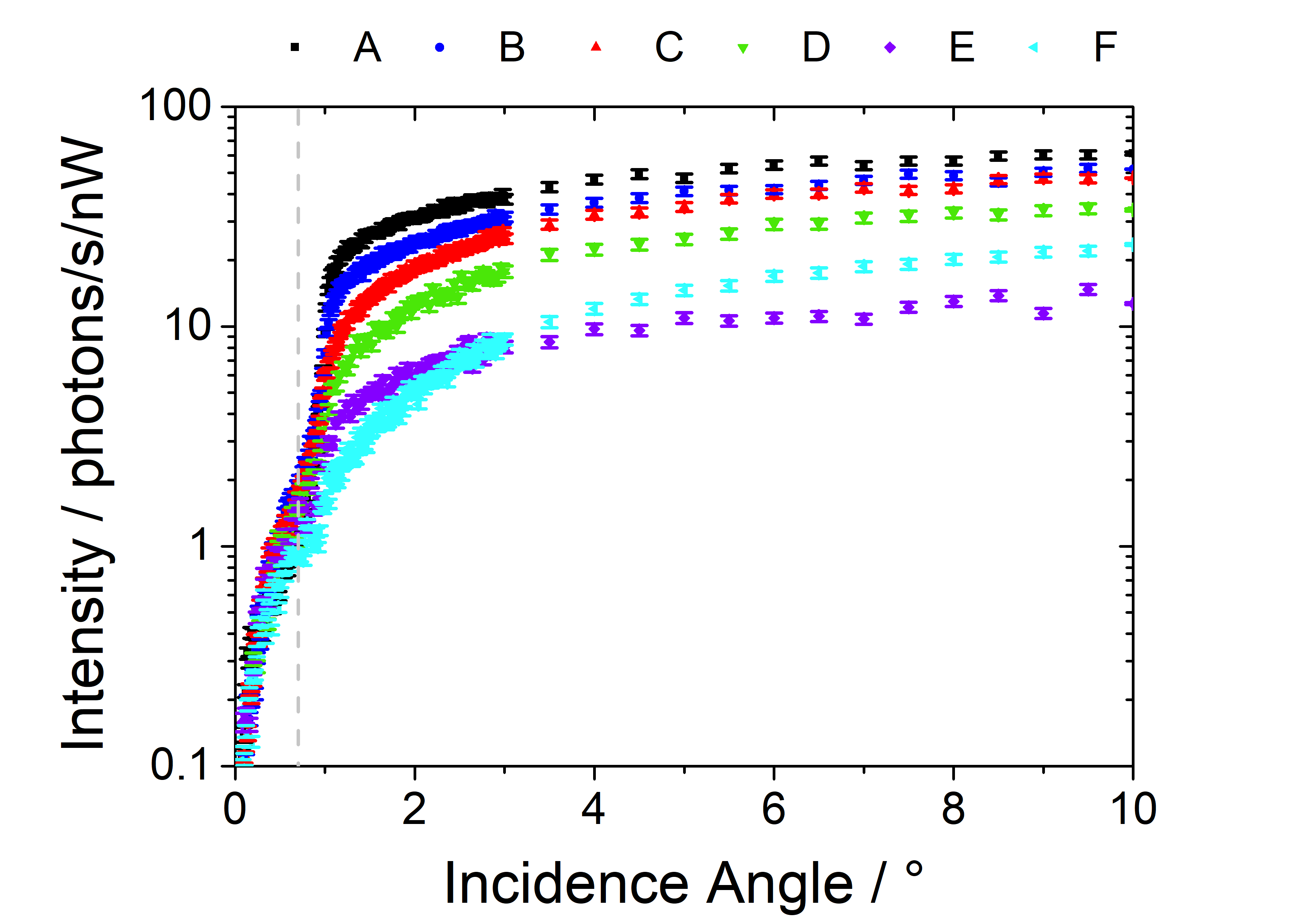}
	\end{center} 
	\caption{The detected Si XRF fluorescence originating from the L-shell for the same samples than displayed in Figures 2 and 4 indicate the growing impact on the attenuation of X-rays within the collected PM. The vertical bar indicates the position typically selected for a TXRF measurement for a Si substrate and the incident photon energy used. In contrast to the reflectivity (Fig. 4) which can be used as an indicator for exceedingly high mass deposition with respect to a reliable quantification of the mass deposition by means of TXRF, scattering or fluorescence signals from the bulk substrate can be used more conveniently as monitors during GIXRF measurements in the range above the critical angle of total external reflection. In both cases, reflectivity and XRF from the bulk substrate, a simple cross-comparison to a blank substrate will be sufficient.}
	\label{fig:figure7}
\end{figure}

\cleardoublepage